\documentclass[aps,prl,twocolumn,amsmath,amssymb,superscriptaddress,hidelinks]{revtex4-2} 

\usepackage{bm}
\usepackage{verbatim} 

\usepackage{graphicx}
\usepackage[usenames,dvipsnames]{color}
\usepackage[colorlinks]{hyperref}
\hypersetup{
  colorlinks,
  citecolor=blue,
  linkcolor=blue,
  urlcolor=blue}

\newcommand{\be}{\begin{equation}}
\newcommand{\ee}{\end{equation}}
\newcommand{\bea}{\begin{eqnarray}}
\newcommand{\eea}{\end{eqnarray}}

\newcommand{\p}{\partial}

\newcommand{\la}{\langle}
\newcommand{\ra}{\rangle}

\newcommand{\sgn}{{\rm sgn}\,}

\renewcommand{\vec}[1]{{\bf #1}}

\newcommand{\dbar}{\raisebox{-0.1ex}[\height][0pt]{$\mathchar'26$}\mkern-12mu {d}}

\begin{document}

\title{Anti-screening and nonequilibrium layer electric phases in graphene multilayers}

\author{Ying Xiong}
\affiliation{Division of Physics and Applied Physics, School of Physical and Mathematical Sciences, Nanyang Technological University, Singapore 637371}
\author{Mark S. Rudner}
\affiliation{Department of Physics, University of Washington, Seattle WA 98195, USA}
\author{Justin C.W. Song}
\email{justinsong@ntu.edu.sg}
\affiliation{Division of Physics and Applied Physics, School of Physical and Mathematical Sciences, Nanyang Technological University, Singapore 637371}

\begin{abstract}
  Screening is a ubiquitous phenomenon through which the polarization of bound or mobile charges tends to reduce the strengths of electric fields inside materials.  Here we show how photoexcitation can be used as a knob to transform conventional out-of-plane screening into {\it anti-screening} -- the {\it amplification of electric fields} -- in multilayer graphene stacks. We find that, by varying the photoexcitation intensity, multiple nonequilibrium screening regimes can be accessed, including near-zero screening, anti-screening, or over-screening (reversing electric fields). Strikingly, at modest continuous wave photoexcitation intensities, the nonequilibrium polarization states become {\it multistable}, hosting light-induced ferroelectric-like steady states with nonvanishing out-of-plane polarization (and band gaps) even in the absence of an externally applied displacement field in nominally inversion symmetric stacks. This rich phenomenology reveals a novel paradigm of dynamical quantum matter that we expect will enable a variety of nonequilibrium broken symmetry phases. 
\end{abstract}

\maketitle

Due to their atomically thin nature, the electronic properties of 
layered quantum materials can be
highly sensitive to out-of-plane electric fields. 
In particular, for multilayer graphene stacks with inversion symmetry (e.g., rhombohedral odd layer or Bernal even layer samples), such fields 
explicitly break
the symmetry and qualitatively alter the layer spatial configuration of 
the material's Bloch eigenstates.
Layer polarization directly opens energy gaps~\cite{Zhang09,Heinz11}, modifies density of states, and turns on quantum geometric electronic responses~\cite{Shimazaki15, Sui15, Deshmukh22}.
As a result, a 
variety of exotic phenomena in layered materials, ranging from ferromagnetism~\cite{Zhou21a, Berrera22, Liu23, Chen23} to superconductivity~\cite{Zhou21b, Zhou22, Zhang23} and Chern insulators~\cite{Han23, Lu23} to optical responses~\cite{Ju17, yin22, ma22, Yang22, song23} are activated by 
layer polarization. These render proximal electrostatic gates a key ingredient for realizing these widely sought phenomena. 

\begin{figure}[t]
\includegraphics[scale=0.22]{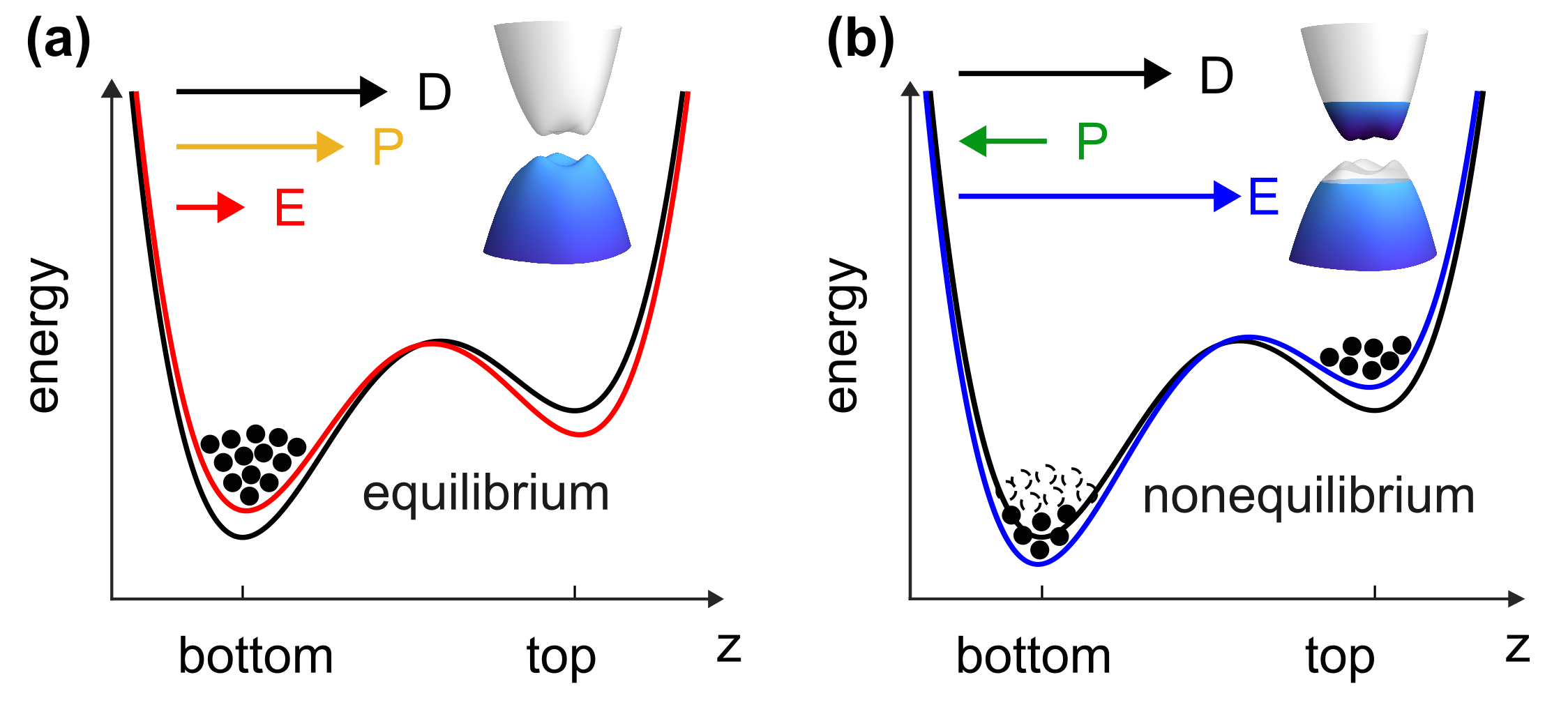}
\caption{{\bf Nonequilibrium anti-screening.} The out-of-plane ($z$-direction) screening behavior of multilayer graphene is intertwined with its layer polarizable band structure. 
  {\bf (a)} When a displacement field $D$ is applied (black arrow), its bands become layer polarized. 
  In equilibrium, electrons 
  occupy 
  states (inset) with lower electrostatic energy (black curve).
  The resulting electric polarization $P$ (orange arrow) 
  screens the displacement field, yielding a reduced
  electric field and
  screened electrostatic potential 
  (red arrow and curve).
  {\bf (b)} Photoexcitation 
  populates states in layers with {\it higher} electrostatic energy.
  For modest intensity, the net nonequilibrium electric polarization 
  can be {\it reversed} relative to the direction of $D$ (green arrow).
  This produces {\it anti-screening}, with polarization {\it amplifying} electric fields (blue arrow and potential curve).} 
\label{fig0}
\end{figure}

Here we show that when interactions combine with nonequilibrium photoexcitation in multilayer graphene, layer polarization can become {\it dynamical}.
Dynamical layer polarization 
produces a rich phenomenology that includes anti-screening (polarization {\it enhances} electric fields, Fig.~\ref{fig0}) as well as overscreening (polarization {\it inverts} electric fields). 
Strikingly, we find nonequilibrium layer polarization can exhibit {\it multistability}, 
with non-equilibrium polarization states 
featuring a 
persistent and hysteretic ferroelectric-like polarization even for vanishing external displacement field. 
We term these ``layer electric phases,'' in analogy with ferroelectrics. 
This persistence is remarkable given that 
the setup is inversion symmetric with no 
out-of-plane electric polarization in equilibrium. 
The protocol we discuss here presents a nonequilibrium paradigm for electronic ferroelectricity sustained by the photoexcitation of its layer polarizable 
electronic bands.
Importantly, this phenomenology arises for incoherent photoexcitation, in contrast to the coherent driving 
required for Floquet engineering~\cite{Rudner20, basov17, oka19}.

We predict anti-screening and layer electric phases can be realized in rhombohedral trilayer graphene (RTG) stacks for moderate light intensities readily accessible in graphene devices~\cite{gabor2011}. 
Relying principally on layer 
polarizable bands, we anticipate layer electric phases also manifests in other layered materials 
with field tunable band structures. 
In RTG, we find sizeable self-sustained layer 
polarization induced band gaps, 
which may open the way to other novel nonequilibrium symmetry-broken electronic phases, e.g., analogous to the displacement field induced ferromagnetism that arises in equilibrium~\cite{Zhou21a, Berrera22, Liu23, Chen23}. 

{\it Electronic layer polarization and hot spots.} We begin by examining 
an inversion symmetric multilayer graphene stack, 
e.g., even (odd) layer Bernal (rhombohedral) graphene.
An electric potential $\Delta/e$ 
sustained across the layers 
breaks inversion symmetry, as captured by the Bloch Hamiltonian~\cite{McCann06, Guinea06, Koshino09}:  
\be
\label{eq:H}
\mathcal{H} (\vec k) = H_0 (\vec k) +  \Delta\mathcal{P}, \quad  \mathcal{P} =  \sum_{\ell,\alpha} (\ell/\ell_0)  |\phi_{\ell,\alpha} \ra \la \phi_{\ell,\alpha} |, 
\ee
where $H_0(\vec k)$ is the unperturbed Bloch Hamiltonian of the inversion symmetric system, 
$\vec k$ is the electron wave vector, 
$\mathcal{P}$ is the layer polarization operator, $e<0$ is the electron charge, and
$|\phi_{\ell,\alpha} \ra$ 
denotes the $\alpha = \{A,B\}$ orbitals 
  on the layer at height $\ell = (\ell_0/2) \{-1, 0, 1\}$ measured from the center of the stack. While our theory is general for multilayer stacks~\cite{McCann06, Guinea06, Koshino09}, 
  for concreteness we focus on RTG~\cite{Heinz11,Guinea06, Koshino09} where
  $\ell_0 = \ell_{\rm top}- \ell_{\rm bottom} = 0.7$ nm. 
  See Supplemental Material ({\bf SM}) for $H_0(\vec k)$ for RTG.

\begin{figure}
\includegraphics[scale=0.55]{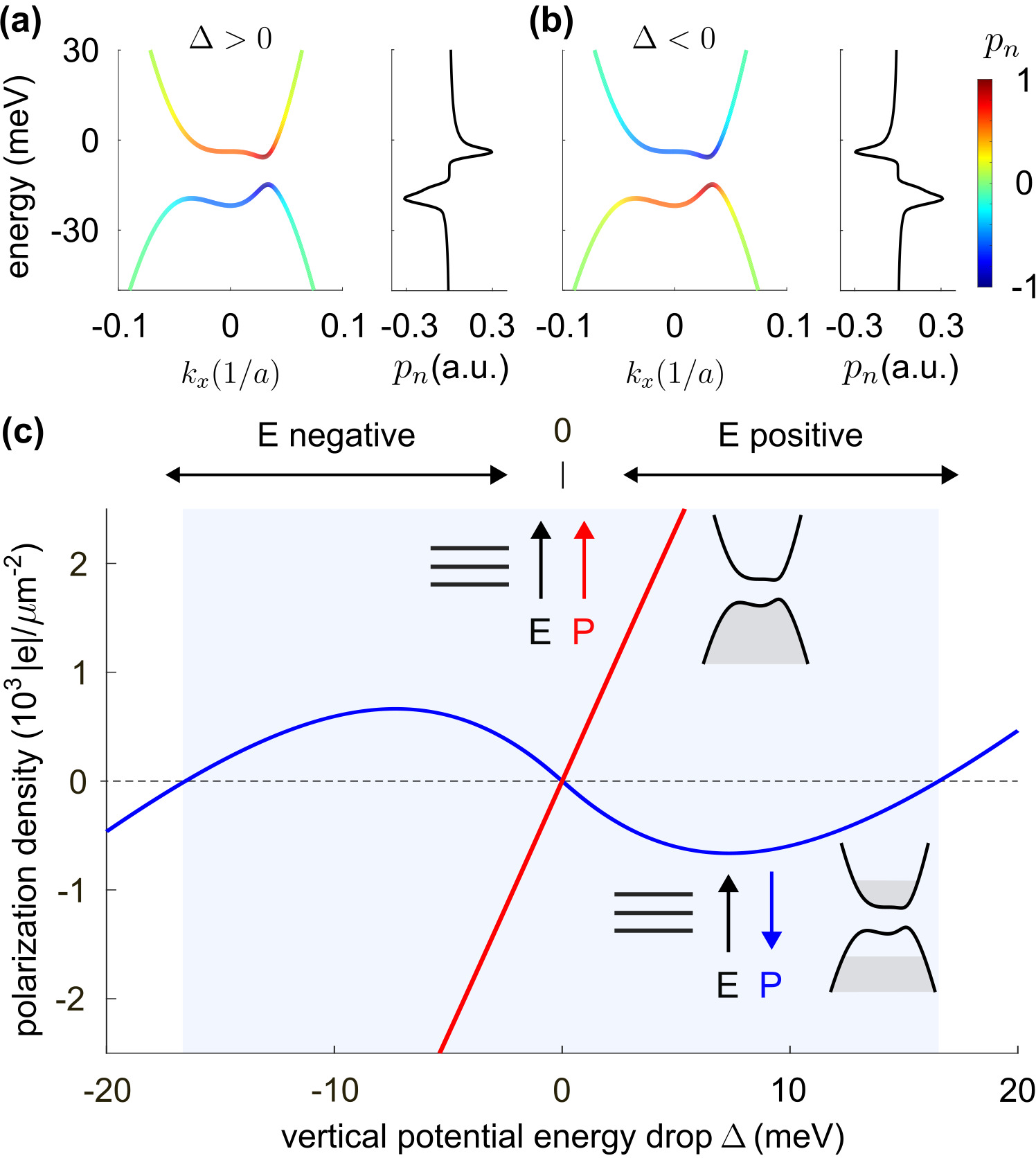}
\caption{{\bf Nonequilibrium layer polarization.} Layer electric polarization density, $P$, in a van der Waals stack depends on the electric potential sustained across the layers. (a,b) The Bloch states (shown for RTG) 
  are most 
  polarized in hot spots near the band edges. 
  The black curves 
  show $p_n (\vec{k}; \Delta)$ integrated over constant energy contours; 
  the peaks/dips illustrate the polarization hot spots. 
  (c) Polarization density at equilibrium (red), with $E$ and $P$ aligned.
  Here $\Delta = -e E \ell_0$ is the 
  total vertical potential energy drop across the stack.
  With photoexcitation (blue), polarization density becomes {\it anti-aligned} with E (blue shaded region).
  Here we used $n_{\rm pe} = 1.5\times 10^{12}\; {\rm cm^{-2}}$, Eq.~(\ref{eq:polarization}) for $P$ at $T_{\rm el} = 60\, {\rm K}$ and $\mathcal{H}(\vec k)$ in Eq.~(\ref{eq:H}).}
\label{figpolarization}
\end{figure}

  The energies $\varepsilon_{n} (\vec{k}; \Delta)$ 
  and Bloch eigenstates $|u_{n} (\vec{k}; \Delta) \ra$ 
  in each band $n$ are highly sensitive to $\Delta$.
  For example, 
  the spectrum at $\Delta = 0$ is gapless due to time-reversal and inversion symmetry.
  Similarly, the Bloch states $|u_{n} (\vec {k}; \Delta = 0) \ra$ are layer {\it unpolarized}.
  Non-zero $\Delta$ breaks inversion symmetry, gapping the energy spectrum; orbitals from the top and bottom layers acquire different electrostatic potentials, leading to layer polarized bands~\cite{McCann06, Guinea06, Koshino09}. 
  The layer polarization $p_n(\vec{k}; \Delta) = \la u_{n} (\vec{k}; \Delta) | \mathcal{P} | u_{n} (\vec{k}; \Delta) \ra$ of each Bloch state is given by 
\be
p_n(\vec{k}; \Delta) = \p \varepsilon_{n} (\vec{k}; \Delta)/\p \Delta,  
\label{eq:layerpolarization} 
\ee
where we used Eq.~(\ref{eq:H}) and the Hellmann-Feynman theorem.
Finite $\Delta$ affects the 
states 
closest to the neutrality point the most, pushing the conduction band, $c$ (valence band, $v$) to higher (lower) energies.
Hence, $p_n$ is concentrated in polarization hot spots close to the band edges (see Fig.~\ref{figpolarization}a,b), with opposite signs for the $c,v$ bands.

Importantly, due to the polarization hot spots, 
electrons close to the $c,v$ band edges 
significantly contribute to the out-of-plane 
electric polarization density $P$~\cite{McCann06, Koshino09, McCannKoshino}, 
\be
P= e \sum_n \int \dbar \vec k \, 
p_n(\vec{k}; \Delta)  f_n (\vec k), 
\label{eq:polarization}
\ee
where $\int \dbar \vec k = \int d^2 \vec k / (2\pi)^2$ and $f_n(\vec k)$ is the distribution function of band $n$. For example, notice that $p_v$ is negative (positive) for $\Delta>0$ ($\Delta <0$) [see Fig.~\ref{figpolarization}a,b].
For a chemical potential at charge neutrality (at equilibrium), $f_n (\vec k) = f^{\rm eq} $, 
the filled valence band yields $P$ aligned with the total electric field and strong out-of-plane dielectric screening (Fig.~\ref{figpolarization}(c), red curve).
Indeed, at low temperatures RTG has a large equilibrium electric susceptibility $\chi_{\rm eq} = P_{\rm eq}/(\epsilon_0 E)  \approx 10 $. 
Here $\epsilon_0$ is the vacuum permittivity and $E$ is the out-of-plane electric field. Throughout we will focus on fields in the out-of-plane direction, omitting the $z$ index for brevity.

{\it Nonequilibrium polarization \& anti-screening.}
When photoexcited out of equilibrium~\cite{DiXiao20,Oles23}, the stack's polarization properties can be dramatically altered. 
Interband photoexcitation combined with rapid thermalization and slow electron-hole recombination in gapped multilayer graphene readily produces an electron-hole liquid at the $c,v$ band edges~\cite{hoffman, yin22, frank23}.
We model 
the thermalized non-equilibrium distribution 
as~\cite{glazman}
\be
f_{c}^{\rm neq} (\vec k) = \frac{1}{1+e^{\beta(\varepsilon_n (\vec k) - \mu_c)}}, \quad f_v^{\rm neq} (\vec k) = \frac{1}{1+e^{\beta (\varepsilon_n (\vec k) - \mu_v)}},
\label{eq:neqdistribution}
\ee
where $f_{c,v}^{\rm neq} (\vec k) $ is the Fermi-Dirac distribution in the conduction and valence bands with quasi-chemical potentials $\mu_c \neq \mu_v$, and $\beta= 1/k_B T_{\rm el}$ denotes the inverse temperature of the electronic system.
These parameters characterize the degree to which the electronic system is pushed out of equilibrium; their values can be obtained self-consistently for a specified photoexcitation intensity, see full account of balance equations below. In what follows, we preserve charge neutrality so that the densities in the conduction and valence bands satisfy $n^{\rm e}_{\rm pe} = n^{\rm h}_{\rm pe} = n_{\rm pe}$ with $n^{\rm e}_{\rm pe}= \int \dbar \vec k  f_c^{\rm neq} (\vec k)$ and $n^{\rm h}_{\rm pe} = \int \dbar \vec k  [1 - f_v^{\rm neq} (\vec k)]$.

To clearly illustrate how polarization is altered out-of-equilibrium, we will first work with fixed gap $\Delta$ (i.e. fixed \emph{total} field across the layers, see below for a self-consistent treatment at fixed \emph{applied} field).
The excited carriers 
induce a change in the electric polarization from its equilibrium value, $\delta P = P^{\rm neq} - P^{\rm eq}$: 
\be
\delta P = e \int \dbar \vec k  \big\{p_c(\vec{k}; \Delta) [ \delta f_c (\vec k)] +  p_v(\vec{k};\Delta) [\delta f_v(\vec k)] \big\}, 
\label{eq:deltaP}
\ee
where $\delta f_{c,v} (\vec k) =  f_{c,v}^{\rm neq} (\vec k) - f^{\rm eq}$.
At charge neutrality, we have $\delta f_{c} >0$ while $\delta f_{v} <0$.
Further, at the band edges, $p_{c}(\vec{k};\Delta)$ and $p_{v}(\vec{k};\Delta)$ have opposite signs.
As a result, the photoinduced change $\delta P$ has a sign {\it opposite} to $P^{\rm eq}$.
Since polarization is concentrated in hot spots at the $c,v$ band edges, as $n_{\rm pe}$ increases, the (opposite signed) $\delta P$ rapidly outweighs $P^{\rm eq}$ from the filled valence bands, yielding a net 
$P^{\rm neq}$ 
oriented {\it against} $E = -\Delta /(e\ell_0)$. 

To illustrate 
this, in Fig.~\ref{figpolarization} we plot the non-equilibrium $P^{\rm neq}$ for a moderate value of $n_{\rm pe} = 1.5 \times 10^{\rm 12} \;{\rm cm^{-2}}$ (blue curve), readily achieved with modest light intensities.
We computed $P^{\rm neq}$ 
numerically 
by substituting Eq.~(\ref{eq:neqdistribution}) into Eq.~(\ref{eq:polarization}) for the Hamiltonian in Eq.~(\ref{eq:H}), for each value of $\Delta$ ($x$-axis), see also {\bf SM}.
This produces $P^{\rm neq}$ 
that is nonlinear (see below) and {\it anti-aligned} to the electric field over a wide window of $\Delta$ values.

\begin{figure}[t!]
\includegraphics[scale=0.55]{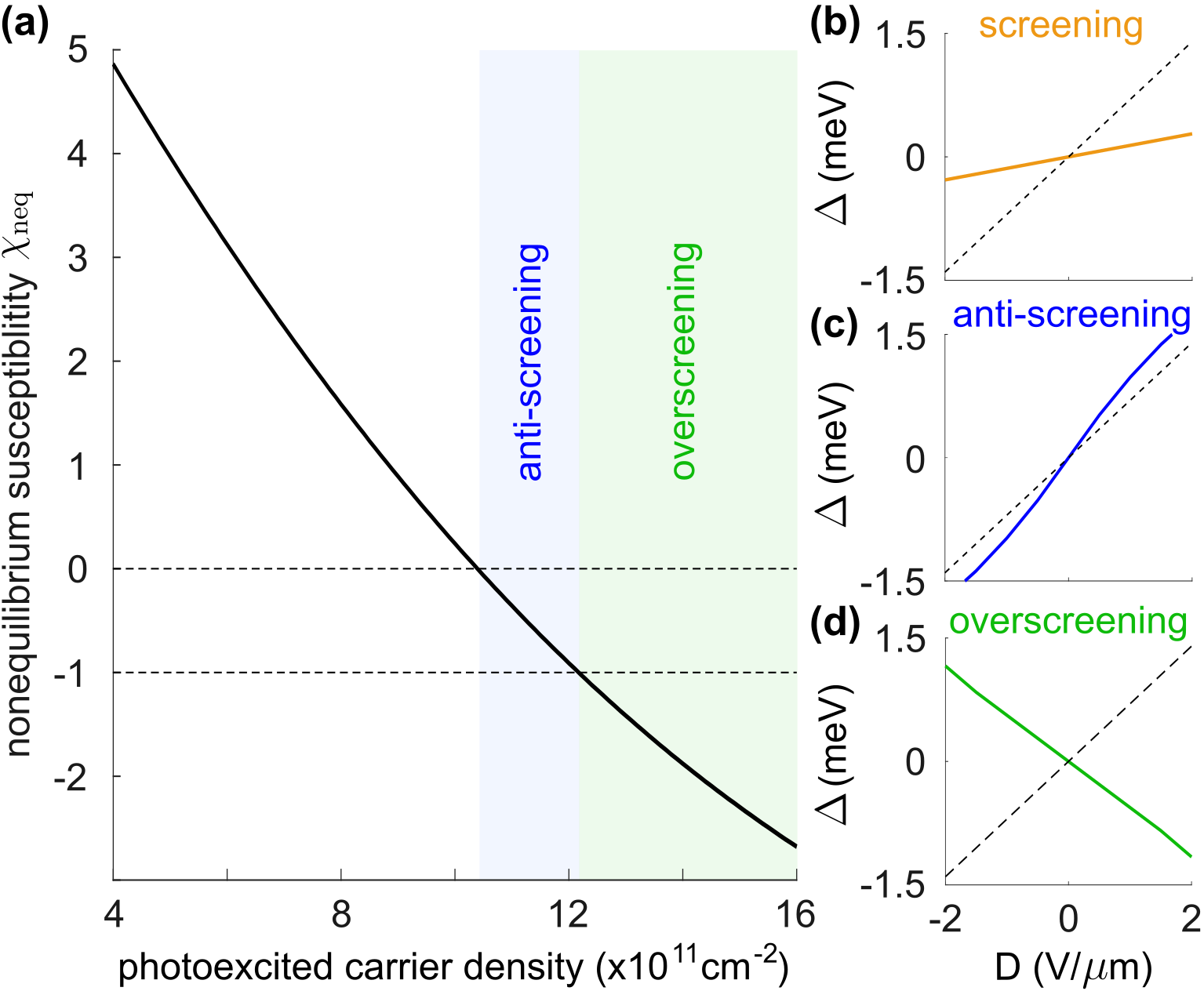}
\caption{{\bf Nonequilibrium electric susceptibility.} (a) The electric susceptibility 
changes from positive to negative as $n_{\rm pe}$ increases. 
(b-d) Self-consistent 
total vertical potential energy drop $\Delta$
as a function of applied $D$, for fixed photoexcited electron densities (b) $n_{\rm pe}=5\times 10^{11} \; {\rm cm^{-2}}$ (orange), (c) $n_{\rm pe}=1.1\times 10^{12} \; {\rm cm^{-2}}$ (blue), (d) and $n_{\rm pe}=1.5 \times 10^{12}\; {\rm cm^{-2}}$ (green) at $T_{\rm el} = 60$ K. Dashed lines denote the potential energy drop in vacuum, $\Delta =  -e D \ell_0/\epsilon_0$; note $e<0$.}
\label{fig3}
\end{figure}

To 
elucidate the behavior of $P^{\rm neq}$, we first examine 
the linear 
regime around 
$\Delta \sim 0$ via 
the nonequilibrium susceptibility
\be
\chi_{\rm neq} = \frac{1}{\epsilon_0}\frac{\partial P^{\rm neq}}{\partial E}=-\frac{e^2 \ell_0}{\epsilon_0} \sum_{n} \int \dbar \vec k  \frac{\partial p_n(\Delta,\vec k)}{\partial \Delta} \Bigg|_{\Delta = 0}\hspace{-2mm} f_n^{\rm neq} (\vec k).
\label{eq:susceptibility}
\ee
In Fig.~\ref{fig3}a we plot $\chi_{\rm neq}$, 
obtained numerically by substituting Eq.~(\ref{eq:neqdistribution}) into Eq.~(\ref{eq:susceptibility}) for the states in Eq.~(\ref{eq:H}). 
For a weak photoexcitation, $\chi_{\rm neq}$ remains positive with too few 
photoexcited electrons and holes to compensate for the dielectric screening of the filled valence bands. 
In this regime, $P^{\rm neq}$ remains aligned with $E$ (white region). 

As $n_{\rm pe}$ increases, $\chi_{\rm neq}$ rapidly decreases, eventually flipping to $\chi_{\rm neq} <0$ (see the blue and green regions in Fig.~\ref{fig3}a). 
This negative susceptibility describes $P$ anti-aligned with $E$. Two regimes are apparent: $-1 < \chi_{\rm neq} < 0$ (blue) and $\chi_{\rm neq} < -1$ (green). As we will now see, these drive anomalous screening behavior. 

Anomalous screening can be described via an electrostatic self-consistency equation $D = \epsilon_0 E + P^{\rm neq}$~\cite{McCann06, Koshino09, McCannKoshino} with a polarization density from Eq.~(\ref{eq:polarization}): 
\be
  \label{eq:self-consistent}
  D + \epsilon_0 \Delta/(e\ell_0) = e \sum_{n} \int \dbar \vec k \, p_n (\vec{k};\Delta) f_n (\vec k),  
\ee
where $D$ is the displacement field.
In Figs.~\ref{fig3}b-d, we show solutions of Eq.~(\ref{eq:self-consistent}) for small $\Delta$, 
illustrating how the total vertical potential drop (energy gap) depends on applied $D$.
For $\chi_{\rm neq} >0$, the polarization density $P^{\rm neq}$ sustained across the layers results in a vertical potential energy drop $\Delta$ (orange, Fig.~\ref{fig3}b) that is smaller than the potential energy drop in vacuum ($-eD\ell_0/\epsilon_0$, dashed); this is conventional screening.
However, when $-1<\chi_{\rm neq} < 0$, $P^{\rm neq}$ acts to amplify $\Delta$ (blue, Fig.~\ref{fig3}c) above the value that would be sustained by $D$ alone in vacuum.
We term this behavior 
{\it anti-screening}. 
Note such amplification behavior has no analog in equilibrium systems~\cite{GiovanniBook}.
  Instead, anti-screening is a hallmark of the nonequilibrium anti-aligned polarization, sustained by the photoexcitation. 
At even larger $n_{\rm pe}$, $\chi_{\rm neq}<-1$ produces {\it over-screening}: $P^{\rm neq}$ overwhelms $D$ yielding values of $\Delta$ (green solid) opposite to $D$ (green, Fig.~\ref{fig3}d). 

Beyond the small $\Delta$ regime,
  $P^{\rm neq}$ exhibits a characteristic 
  non-monotonic dependence on $\Delta$ (blue, Fig.~\ref{figpolarization}c). 
  At small $\Delta$ and sufficiently large $n_{\rm pe}$, the photoinduced anti-aligned polarization contribution $\delta P$ in Eq.~(\ref{eq:deltaP}) dominates over the conventional screening of the filled valence bands. 
 For large $\Delta$, the anti-aligned contribution of the photoexcited carriers saturates while the screening by the valence bands continues to increase, ultimately restoring conventional screening.

{\it Layer electric phases.}
As we will now see, in the overscreening regime this nonlinearity 
leads to {\it multiple stable solutions} of the vertical potential drop $\Delta$ (and polarization $P^{\rm neq}$) for fixed 
$D$ in Eq.~(\ref{eq:self-consistent}).
Strikingly, finite vertical potential drops $\Delta \neq 0$ persist even when $D=0$, thus realizing 
a nonequilibrium ferroelectric-like broken inversion symmetry activated by photoexcitation. 

\begin{figure}
\includegraphics[scale=0.58]{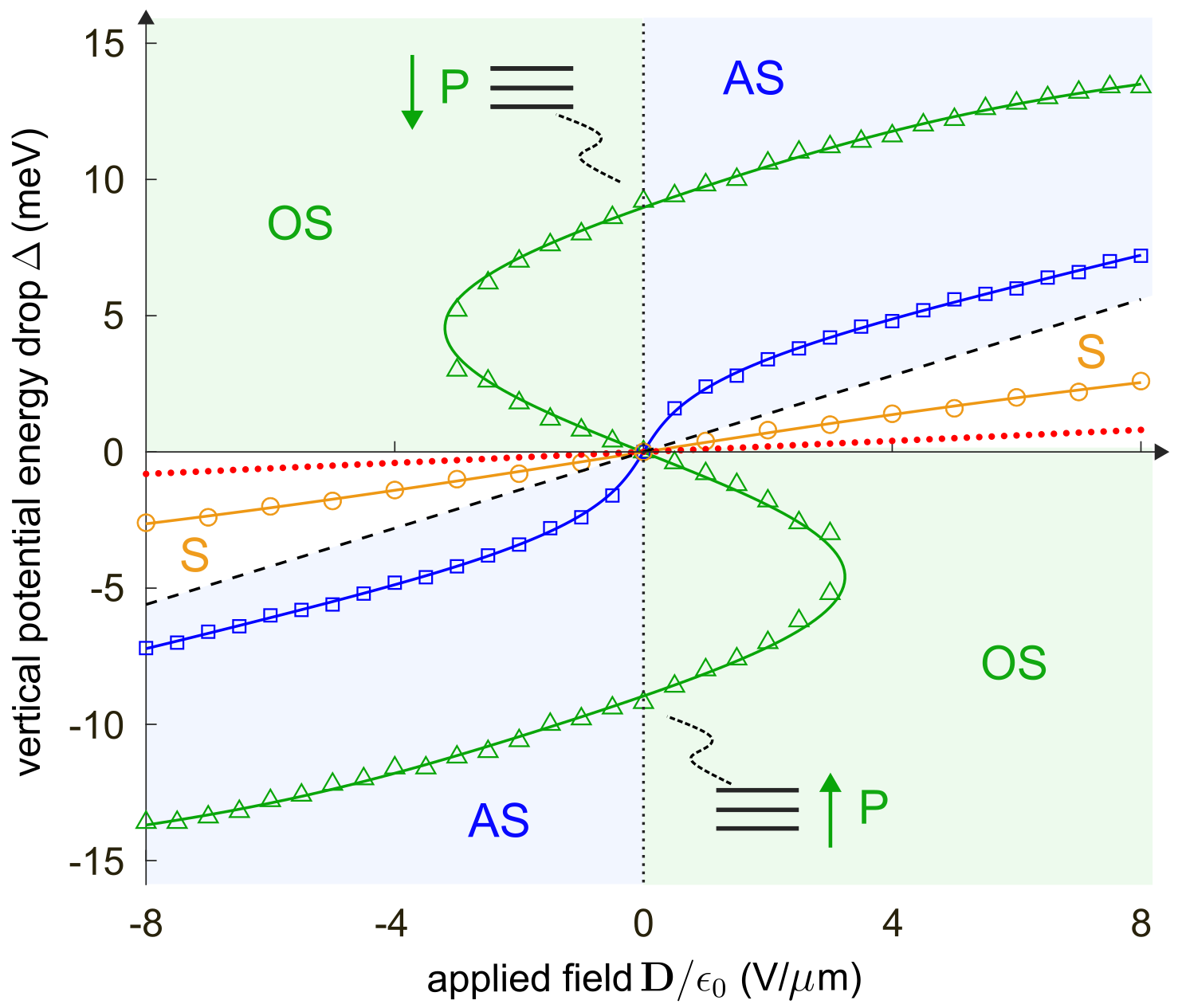}
\caption{{\bf Self-consistent layer electric phases.} Self consistent 
  total vertical potential energy drop across the stack (energy gap) $\Delta$ for light intensities $5 \; {\rm kW/cm^2}$ (orange), $14\; {\rm kW/cm^2}$ (blue) and $25 \; {\rm kW/cm^2}$ (green), found by solving Eq.~(\ref{eq:self-consistent}-\ref{eq:balance}). The open circles, squares and triangles are the numerical self-consistent solutions; the solid lines are guides to the eye.
  The red solid dots 
  show the screened gap $\Delta$ at equilibrium; the black dashed line denotes the potential energy drop in vacuum, $\Delta =  -e D \ell_0/\epsilon_0 $. For small intensity (orange curve, corresponding to the white region in Fig.~\ref{fig3}), the vertical potential energy drop $\Delta$ is almost linear in $D$, similar to that in an equilibrium system. As intensity increases, $\Delta$ becomes highly nonlinear. For large enough intensities (the green region in Fig.~\ref{fig3}), multiple self consistent solutions manifest: the system sustains an inversion symmetry breaking layer electric phase with a nonzero $\Delta$ even at $D =0 $. Here we used $\hbar \omega = 50 \, {\rm meV}$. }
\label{fig4}
\end{figure} 

To track this unconventional behavior systematically, we explicitly analyze the balance equations that determine the non-equilibrium distributions in Eq.~(\ref{eq:neqdistribution}) for a given incident light field (in-plane) $\vec E_{\rm light}(t) = (1/2) \vec E_0 e^{i\omega t} + c.c.$; here $\vec E_{\rm light}(t)$ produces interband transitions with rate $W$.  The quasi-chemical potential $\mu_{c,v}$ and electronic temperature $T_{\rm el}$ in Eq.~(\ref{eq:neqdistribution}) can be tracked by accounting for the photoexcited carrier density $n_{\rm pe}$ as well as the electronic heat density $Q$~\cite{glazman}:
\be
  \frac{d n_{\rm pe}}{dt} = W - R (n_{\rm pe}), \quad \frac{dQ}{dt} =  W \delta \varepsilon - J_{\rm e-ph} (T_{\rm el}, T_{\rm ph}),
  \label{eq:balance}
\ee
where $R$ is the electron-hole recombination rate~\cite{rana,hoffman,haug,frank23}, $J_{\rm e-ph}$ is the electron-phonon cooling rate~\cite{macdonald09,dassarma09}, and $T_{\rm ph}$ is the lattice temperature. Here $\delta \varepsilon$ is the energy each photoexcited electron-hole pair injects into the system above $\mu_{c,v}$~\cite{glazman}. 
A full account of recombination rates used (estimated from experiments in gapped multilayer graphene) as well as electron-phonon cooling, can be found in the {\bf SM}. At steady state, $dn_{\rm pe}/dt = 0$ and $dQ/dt = 0$. We take a conservative estimate for $\delta \varepsilon \approx \hbar\omega - (\mu_c-\mu_v)$ for $\hbar \omega > \mu_c - \mu_v$, 
by assuming all of the electron's kinetic energy is rapidly thermalized into the electronic system as heat~\cite{footnote1, song15review}.

We obtain self-consistent solutions for $\Delta$ and $n_{\rm pe}$ 
by solving Eqs.~(\ref{eq:neqdistribution}), (\ref{eq:self-consistent}), and Eq.~(\ref{eq:balance}) together.
In Fig.~\ref{fig4} we show numerically obtained 
self-consistent solutions of $\Delta$ for varying light intensities $I$ with $T_{\rm ph} = 10\, {\rm K}$; 
 the effective electron temperatures $T_{\rm el}$ obtained ranged from about 45 to 65 K.
As intensity increases (orange to blue to green curves), 
screening (unshaded), anti-screening (blue shaded), and overscreening (green shaded) regimes are accessed sequentially. 
Note that even at moderate intensities (blue), $\Delta$ vs $D$ 
rises sharply above the dashed line, indicating significant amplification.
For comparison, the equilibrium (screened) potential/gap is shown in red.

Most striking is the pronounced S-shaped vertical electric potential drop $\Delta$ vs.~displacement field $D$ obtained at higher intensity 
(green).
For a range of values of $D$ around zero, we find {\it multiple solutions} for $\Delta$ corresponding to distinct self-consistent configurations of polarization for each value of the applied field $D$.
The middle branch corresponds to the overscreened regime,
while the outer 
branches possess self-sustained nonequilibrium polarization $P^{\rm neq}$, even at $D = 0$. 

These layer electric phases can be understood via a feedback process: the photoexcited carriers form dipoles 
with an induced electric field that re-enforces the vertical potential drop $\Delta/e$.
The vertical potential drop $\Delta/e$ in turn 
polarizes the photoexcited carriers, with those in the filled valence bands developing an opposite polarization that stabilizes $\Delta$ at its steady state value (see discussion of Fig.~\ref{figpolarization}(c) above and {\bf SM} for stability analysis).
This 
feedback is key to the layer electric phases, enabling self-sustained gap opening of 
$\sim 10$ meV that persist even at $D \to 0$ for readily achievable light intensities. 

We expect pronounced hysteresis of the vertical potential drop $\Delta/e$ and its associated $P^{\rm neq}$ as $D$ is swept.
The hysteretic window as well as the maximum $\Delta$ values are controlled by light intensity and can reach remnant polarization values of $\Delta P^{\rm neq} = P^{\rm neq}_+ - P^{\rm neq}_- \approx 0.024 \; \mu C {\rm cm}^{-2}$ for $ 25 \; {\rm kW}\, {\rm cm} ^{-2}$ irradiation. Here $\pm$ indicate forward and backward $D$ sweeps. These remnant polarizations are similar in magnitude to the electric polarization recently observed in moir\'{e} heterostructures~\cite{Zeng20} and can be tracked using a proximal graphene layer as an electric field sensor~\cite{Zeng20, Fei18}. 
Lastly, 
  we note that for $D\neq 0$ photoexcitation activates giant enhancements to induced gaps 
  (see red dots vs green triangles in Fig.~\ref{fig4}).
  These
  signatures of the anomalous screening regimes 
  may also be accessed in short time scale pump/probe experiments. 

Photoexcitation activates a strong nonlinear coupling between electric polarization and band structure to transform a ground state dielectric into a nonequilibrium ``layer electric.'' 
While we focused on multilayer graphene stacks due to their out-of-plane field sensitive bands, a variety of layered materials (e.g., MnBi$_2$Te$_4$~\cite{Gao2021}) also possess layer- and spin-polarization dependent electronic structures, rendering them promising targets for other nonequilibrium states of matter~\cite{RudnerSong19}. Realized with incoherent drives, layer electric phases open a rich phase space of nonequilibrium phenomena to explore, including novel symmetry broken phases, spatial pattern formation, and unconventional domain dynamics.

{\it Acknowledgements.} We gratefully acknowledge useful conversations with Qiong Ma, Roshan Krishna Kumar, Giovanni Vignale, Leonid Glazman, Brian Skinner, and Erez Berg. This work was supported by the Singapore Ministry of Education Tier 2 grant MOE-T2EP50222-0011. M.R.~acknowledges the Brown Investigator Award, a program of the Brown Science Foundation, the University of Washington College of Arts and Sciences, and the Kenneth K.~Young Memorial Professorship for support.



\clearpage

\newpage

\setcounter{equation}{0}
\setcounter{figure}{0}
\renewcommand{\theequation}{S\arabic{equation}}
\renewcommand{\thefigure}{S\arabic{figure}}

\renewcommand{\bibnumfmt}[1]{(#1)}

\onecolumngrid

\begin{center}
\textbf{\large Supplementary Material for ``Anti-screening and nonequilibrium layer electric phases in graphene multilayers"} 
\end{center}

\section{Tight-binding model for Rhombohedral trilayer graphene}

In the main text, we illustrate how nonequilibrium photoexcitation induces anomalous screening behavior 
and layer electric phases in multilayer graphene.
For illustration, we use rhombohedral trilayer graphene (RTG) as our primary example. In this subsection, we describe the tight-binding Hamiltonian for RTG used in our numerical calculations.
First note that RTG is inversion symmetric in the absence of applied fields.
When an out-of-plane electric field is sustained across its layers, inversion symmetry is broken and a band gap is opened.

At the single-particle (or mean-field) level, the electrons can be described by a tight-binding Hamiltonian 
\be\label{eq:HSupp}
\mathcal H (\vec k) = H_0 (\vec k) + \Delta \mathcal P,
\ee
where $\Delta$ is the total potential energy drop from layer $l = \ell_0/2$ to layer $l = -\ell_0/2$ (i.e., between the top and bottom layers), $\mathcal P$ is the layer polarization operator, and $H_0 (\vec k)$ is the Hamiltonian of the unperturbed stack, in the absence of any applied fields.

In the basis of ${\rm (A_1, \; B_3,\; B_1, \; A_2, \; B_2, \; A_3)}$, where $A_i$ and $B_i$ denote orbitals on the $A$ and $B$ sublattices of the graphene sheet in layer $i$, respectively, $H_0(\vec k)$ is given by~\cite{Zhou21a}
\be \label{eq:H_pristine}
H_0 (\vec k) = \begin{pmatrix}
 \Delta_2 +\delta & \frac{1}{2}\gamma_2 & -\gamma_0 t(\vec k) & -\gamma_4 t(\vec k) & -\gamma_3 t(\vec k)^* & 0 \\
	\frac{1}{2} \gamma_2 & \Delta_2 + \delta & 0 & -\gamma_3 t(\vec k) & - \gamma_4 t(\vec k)^* & - \gamma_0 t(\vec k)^* \\
	-\gamma_0 t(\vec k)^* & 0 & \Delta_2 & \gamma_1 & -\gamma_4 t(\vec k) &0 \\
	-\gamma_4 t(\vec k)^* & -\gamma_3 t(\vec k)^* & \gamma_1 & -2 \Delta_2 & -\gamma_0 t(\vec k) & -\gamma_4 t(\vec k) \\
	-\gamma_3 t(\vec k) & -\gamma_4 t(\vec k) & -\gamma_4 t(\vec k)^* & -\gamma_0 t(\vec k)^* & -2\Delta_2 & \gamma_1 \\
	0 & -\gamma_0 t(\vec k) & 0 & -\gamma_4 t(\vec k)^* &\gamma_1 & \Delta_2 
\end{pmatrix} ,
\ee
and the polarization operator is 
\be\label{eq:pz}
\mathcal P = \frac{1}{2} \begin{pmatrix}
1 & 0 & 0 & 0 & 0 & 0 \\
0 & -1& 0& 0 & 0 & 0 \\
0 & 0 & 1 & 0 & 0 & 0 \\
0 & 0 & 0 & 0 & 0 & 0 \\
0 & 0 & 0 & 0 & 0 & 0 \\
0 & 0 & 0 & 0 & 0 & -1  \end{pmatrix}.
\ee
Here $\gamma_0  = 3.1$ eV, $\gamma_1 = 0.38$ eV, $\gamma_2 = -0.015$ eV, $\gamma_3 = -0.29$ eV, and $\gamma_4 = -0.141$ eV are the hopping parameters, $t(\vec k) = e^{i \frac{a}{\sqrt{3}} k_y} + 2 e^{-i \frac{a}{2 \sqrt{3}} k_y} \cos \left( \frac{a}{2} k_x \right)$, $\delta = -0.0105$ eV accounts for the on-site potential energies of orbitals $A_1$ and $B_3$, which are not immediately next to an atom of the middle layer, and $\Delta_2 = -0.0023$ eV accounts for the potential difference between the middle layer and the average of the outer layers~\cite{Zhou21a}.

\section{Dynamics of photoexcited carriers}
In this section, we discuss the time evolution of the photoexcited nonequilibrium charge carrier density in a multilayer graphene stack (e.g., RTG). To do so, we consider a layered system pumped by an incident EM irradiation that induces interband transitions. In what follows, we will fix the system at charge neutrality. Upon photoexcitation of electron-hole pairs, we denote the nonequilibrium electron/hole density as $n_{\rm pe} = n_{\rm pe}^{\rm e} = n_{\rm pe}^{\rm h}$. We describe the dynamics of the nonequilibrium carrier density 
by Eq.~(8) of the main text. 

To model the rate of light induced interband transitions, $W$, we consider the bare Hamiltonian of the multilayer graphene stack (e.g., RTG) as shown 
in Eq.~(\ref{eq:HSupp}). In the presence of incident light, the electronic system couples to the EM field $\vec E_{\rm light}(t) = (1/2) \vec E_0 e^{i \omega t} + c.c.$ via 
\be
H_{\rm int} = e \vec A(t) \cdot \hat{\vec v}, 
\ee
where $\hat{\vec v} = \nabla_{\vec k} \mathcal H(\vec k) /\hbar $ is the velocity operator and $\vec A(t)$ is the vector potential such that $\vec E_{\rm light} (t) = - \partial_t \vec A(t)$. We focus on the frequency regime where incident light (with intensity $I= \epsilon_0 c |\vec{E}_0|^2/2$ and $c$ the speed of light) induces interband transitions from the valence band to the conduction band. The interband transition rate per unit area can be simply captured by Fermi's golden rule: 
\be
W =  \frac{\pi e^2 }{2 \hbar \omega^2 } 
\int \dbar \vec k |\vec E_0 \cdot \vec v_{cv} (\vec k)|^2 [ f_v (\vec k)  - f_c (\vec k )] \delta(\varepsilon_c (\vec k) - \varepsilon_v (\vec k) - \hbar \omega),
\label{eq:interbandrate}
\ee 
where $\vec v_{cv} (\vec k) = \la u_c (\vec k) |\vec{\hat v} | u_v(\vec k)\ra$ is the velocity matrix element, $\varepsilon_n(\vec k)$ is the energy dispersion of band $n$ with corresponding Bloch eigenstates $|u_n(\vec k)\ra$, $f_n (\vec k)$ is the electronic distribution function, and $\delta(\varepsilon_c (\vec k) - \varepsilon_v (\vec k) - \hbar \omega)$ accounts for energy conservation. 
At equilibrium, $f_n (\vec k) = f^{\rm FD} (\varepsilon_n(\vec k), \mu, T)$ is the Fermi-Dirac distribution at chemical potential $\mu$ and temperature $T$. 
For the out-of-equilibrium photoexcitation, as we discuss in the main text, the photoexcited electrons and holes quickly thermalize to band edges and can be described the thermalized distribution functions with quasi-chemical potentials for the valence and conduction bands as shown 
Eq.~(4) in the main text. We numerically calculated Eq.~(\ref{eq:interbandrate}) where in accounting for a discretized $k$-space, we used a Gaussian with phenomenological standard deviation $3\, {\rm meV}$ to approximate the Dirac delta function. In computing the interband transition rates, we used a linearly polarized incident EM field as a simple illustration. 

The electron-hole recombination rate $R (n_{\rm pe})$ depends on the product of electron density and hole density and, in the high-density regime for graphene systems, can be modeled as~\cite{rana,haug}
\be 
R(n_{\rm pe}) = \gamma n_{\rm pe}^2,
\ee
where $\gamma$ is a phenomenological constant. Gapped multilayer graphene systems typically exhibit long carrier lifetimes and small values of $\gamma $ ranging from $0.03 \sim 0.1 \;{\rm cm^2 \; s^{-1}}$, directly estimated from the experimental photoconductivity of gapped multilayer graphene with gaps of order $10\, {\rm meV}$ in Ref.~\cite{frank23}. As a result, they are ideal platforms to achieve large nonequilibrium electron-hole densities. For our numerical calculations, we used $\gamma = 0.05 \; {\rm cm^2 s^{-1}}$ as an illustration within the range of $\gamma$ for gapped multilayer graphene systems experimentally achieved; we note that the self-consistent gaps achieved in the layer electric phase shown in the main text are similar to that found in the gapped multilayer graphene stacks experimentally probed in Ref.~\cite{frank23}.

\section{Cooling of photoexcited carriers}

We describe electron cooling via electron-phonon scattering. Since optical phonons in graphene stacks have much higher energy than the quasi chemical potential, here we focus on the scattering from acoustic phonons as the main cooling channel. As a simple illustration, the rate of heat transfer per unit area from the electronic system to the lattice (assuming intraband scattering through a simple one-phonon channel)~\cite{macdonald09,dassarma09} can be estimated as
\be
J_{\rm e-ph} = \frac{2\pi N}{\hbar} \sum_{n,\vec k} \int \dbar \vec p \, \dbar \vec q 
[\varepsilon_n(\vec k) - \varepsilon_n (\vec p)] g^n_{\vec k\vec p, \vec q}  [f_n (\vec k ) - f_n (\vec p) ] [n_{ph} (\hbar \omega_{\vec q}) - n_e (\hbar \omega_{\vec q})] \delta_{\vec k, \vec p+\vec q} \delta ( \varepsilon_n(\vec k) - \varepsilon_n (\vec p) - \hbar \omega_{\vec q}), 
\ee
where $N=3$ denotes the number of phonon branches, $\vec p$ and $\vec k$ are the initial and final Bloch wave vectors of the electron, $\vec q$ is the phonon wave vector, $\hbar \omega_{\vec q} = s\hbar  |\vec q|$ is the phonon energy with $s$ the acoustic phonon speed, $n_{\rm ph(e)} (\hbar \omega) = 1/(e^{\hbar \omega /k_BT_{\rm ph(e)}}-1) $ is the Bose function with $k_B$ Boltzmann's constant, $ g^n_{\vec k \vec p, \vec q} = \frac{\hbar^2 }{2\rho \hbar \omega_{\vec q}} D^2 q^2 F^n_{\vec k \vec p}$ is the electron-phonon coupling constant with $D$ an effective deformation potential, $ F^n_{ \vec k \vec p} = |\la u_n(\vec k)| u_n (\vec p) \ra|^2$ is the overlap integral, and $\rho$ is the areal mass density. Here $\delta_{\vec k, \vec p+\vec q}$ and $\delta (\varepsilon_n(\vec k) - \varepsilon_n (\vec p) - \omega_{\vec q})$ account for momentum and energy conservation during the scattering process. In the numerical calculations, the Dirac delta function is approximated by a Gaussian with standard deviation of 3 meV, 
similar to the interband photoexcited transition rate (see above). For our simulations, we used $D =30 \, {\rm eV}$, $\rho = 22.8 \times 10^{-7} {\rm kg}/{\rm m}^2$, and $s = 2.6 \times 10^4 \, {\rm m}/{\rm s}$~\cite{macdonald09,dassarma09}. Lastly, in our numerical estimate we have used the same temperature for both electron and hole populations.

\section{Stability analysis of the self consistent steady state solutions}
In this section, we analyze the stability of the self consistent steady state solutions upon a small perturbation in the total vertical potential energy drop (gap size) $\Delta$. For a fixed light irradiation intensity, we consider a steady state solution $(\Delta^{\rm st}, \; n_{\rm pe}^{\rm st}, \; T^{\rm st})$. We examine the time evolution of the system after the perturbation is applied. The steady state solution is stable if it evolves back to itself after a small perturbation, and is unstable if it evolves further away from the original solution. In the following,  
we assume the temperature to be constant since a small variation of $\Delta$ and $n_{\rm pe}$ only leads to negligible change in temperature. 

First, we consider a small perturbation in the gap size $\tilde \Delta = \Delta^{\rm st} + \delta \Delta$. Note that for the perturbed gap size, the self consistency relation in Eq.~(7) of the main text must be satisfied. 
For a given gap size and temperature, the polarization $P$ implicitly depends on the nonequilibrium carrier density via the quasi Fermi energies $\mu_c$ and $\mu_v$, see Eqs.~(3)-(4) in the main text. Thus, the perturbed $\tilde \Delta$ demands a corresponding  $\tilde n_{\rm pe}$ such that the self consistency requirement Eq.~(7) is satisfied. For the perturbed system with $\tilde \Delta$ and $\tilde n_{\rm pe}$, we numerically computed $d\tilde n_{\rm pe}/dt$. If $\sgn (\tilde n_{\rm pe} - n_{\rm pe}^{\rm st}) \cdot \sgn (d\tilde n_{\rm pe}/dt) < 0$, then the steady state solution is stable as the perturbed system will evolve back to its original state (without the small perturbation). 
Indeed, our numerical calculation finds that all three branches of self consistent steady state solution (as shown in Fig.~4 of the main text) possess a $\sgn (\tilde n_{\rm pe} - n_{\rm pe}^{\rm st}) \cdot \sgn (d\tilde n_{\rm pe}/dt)$ (upon small perturbation of $\delta \Delta$) that is negative. As a result, we conclude that the three branches are stable.


\begin{thebibliography}{99}
\bibitem{footnote1} In reality, $\delta \varepsilon$ is likely to be smaller due to other competing relaxation mechanisms in the photoexcitation cascade (e.g., relaxation to lattice degrees of freedom) yielding a lower (and more favorable) $T_{\rm el}$
\end{thebibliography}
\end{document}